%% file: paper.tex
\documentclass[submission,copyright,creativecommons]{eptcs}
\providecommand{\event}{Prole 2014} 
\usepackage{breakurl} 
\usepackage{amssymb}
\usepackage{graphicx}
\usepackage{wrapfig} 

\title{Improving the Deductive System DES with Persistence by Using SQL DBMS's}
\author{Fernando S\'aenz-P\'erez
\institute{Grupo de Programaci\'on Declarativa (GPD)\\
Dept. Ingenier\'{\i}a del Software e Inteligencia Artificial\\
Universidad Complutense de Madrid, Spain}
\email{fernan@sip.ucm.es}
}
\def\titlerunning{Improving DES with Persistence and SQL DBMS's}
\def\authorrunning{F. S\'aenz-P\'erez}

\newcommand{\eval}[2]{\parallel#1\parallel_{#2}} 
\newcommand{\myfontcodesize}{\fontsize{10.5}{11}}

\begin{document}
\maketitle

\begin{abstract}
This work presents how persistent predicates have been included in the in-memory deductive system DES by relying on external SQL database management systems.
We introduce how persistence is supported from a user-point of view and the possible applications the system opens up, as the deductive expressive power is projected to relational databases.
Also, we describe how it is possible to intermix computations of the deductive engine and the external database, explaining its implementation and some optimizations.
Finally, a performance analysis is undertaken, comparing the system with current relational database systems.
\end{abstract}

\input{01} 
\input{03} 
\input{05} 
\input{06} 
\input{07} 
\input{08} 

\bibliographystyle{eptcs}

\end{document}

%% file: 01.tex
\section{Introduction}
\label{sect:intro}

Persistence is one of the key features a database management system (DBMS) must fulfil.
Such features are found in the well-known acronym ACID,
where in particular D stands for Durability (i.e., persistence of data along different user sessions) \cite{silberschatz2010database}.
This way, updates in the database must be persistent in a non-volatile memory, as secondary storage (typically, the file system that the host operating system provides).
Whereas persistence in relational DBMS's are given for granted, deductive databases have been traditionally implemented as in-memory database systems (as, e.g., DLV~\cite{dlv-2006}, XSB~\cite{DBLP:journals/tplp/SwiftW12}, bddbddb~\cite{bddbddb-2005}, Smodels~\cite{DBLP:journals/corr/cs-AI-0003033}, DES~\cite{DBLP:conf/aplas/Saenz-PerezCG11}, \ldots)
Some logic programming systems also allow persistent predicates, as Ciao Prolog does \cite{DBLP:conf/padl/CorreasGCCH04} (but only for the extensional part of the database).


In this work, we present an approach for adding predicate persistence to the deductive system DES (\url{des.sourceforge.net}) \cite{DBLP:conf/aplas/Saenz-PerezCG11} relying on external SQL DBMS's via ODBC bridges.
Enabling persistence leads to several advantages:
1) Persistent predicates with transparent handling, also allowing updates.
Both the extensional (EDB, i.e., facts) and intensional (IDB, i.e., rules) databases can be persistent.
2) Interactively declare and undeclare predicates as persistent.
Applications for this include database migration (cf. Section \ref{sect:migration}).
3) Mix both deductive solving and external SQL solving.
On the one hand, the system takes advantage of the external database performance (in particular, table indexing is not yet provided by DES) and scalability.
On the other hand, queries that are not supported in an external database (as hypothetical queries or recursive queries in some systems) can be solved by the deductive engine.
So, one can use DES as a front-end to an external database and try extended SQL queries that add expressiveness to the external SQL language (cf. Sections \ref{sect:extending-expressivity} and \ref{sect:business-intelligence}).
4) Database interoperability.
As several ODBC connections are allowed at a time, different predicates can be made persistent in different DBMS's, which allows interoperability among external relational engines and the local deductive engine, therefore enabling business intelligence applications (cf. Section \ref{sect:business-intelligence}).
5) Face applications with large amounts of data which do not fit in memory.
Predicates are no longer limited by available memory (consider, for instance, a 32bit OS with scarce memory); instead, persistent predicates are using as much secondary storage as needed and provided by the underlying external database.
Predicate size limit is therefore moved to the external database.

Nonetheless, a few deductive systems also integrated persistence or database connections, as  DLV$^{DB}$ \cite{Terracina:2008:ERQ:1348834.1348835},
MyYapDB \cite{DBLP:conf/jelia/FerreiraR04}, and LDL++~\cite{ldl-2003}.
One point that makes DES different from others is the ability to declare on-the-fly a given predicate as persistent and to drop such a declaration.
This is accomplished by means of assertions, which together with a wide bunch of commands, make this system amenable for rapid experimenting and prototyping.
In addition, since predicates can be understood as relations, and DES enjoys SQL, relational algebra (RA) and Datalog as query languages (SQL and RA are translated into Datalog), a persistent predicate can be used in any language and a given query can mix persistent predicates located at different databases.
Those systems neither support full-fledged duplicates (including rules as duplicate providers), nor null-related operations, nor top-N queries, nor ordering metapredicates, nor several query languages accessing the same database (including Datalog, SQL, and extended relational algebra) as DES does \cite{DBLP:conf/aplas/Saenz-PerezCG11}.
Such features are required for supporting the already available expressiveness of current relational database systems.
In addition, no system support hypothetical queries and views for decision support applications \cite{sae13c-ictai13}.

Organization of this paper proceeds as follows.
Section \ref{sect:persistence} describes our approach to persistence, including in Subsection \ref{sect:intermixing} a description of intermixing query solving as available as the result of embodying external DBMS access into the deductive engine, as well as some optimizations.
Section \ref{sect:applications} lists some applications for which persistence in a deductive system is well-suited.
Next, Section \ref{sect:performance} compares performance of this system w.r.t. DBMS's, and the extra work needed to handle persistent data.
Finally, Section \ref{sect:conclusions} summarizes some conclusions and points out future work.

%% file: 03.tex
\section{Enabling Persistence}
\label{sect:persistence}

For a given predicate to be made persistent in an external SQL database, type information must be provided because SQL is strong-typed.
As DES allows optional types for predicates (which are compatible with those of SQL) 
the system can take advantage of known type information for persistence.
Note that, although the predicate to be made persistent has no type information, it may depends on others that do.
This means that the declared or inferred type information for such a predicate must be consistent with other's types.
To this end, a type consistency check is performed whenever a predicate is to be made persistent.

\subsection{Declaring Persistence}
\label{sect:declaring-persistence}
We propose an assertion as a basic declaration for a persistent predicate, similar to \cite{DBLP:conf/padl/CorreasGCCH04}.
The general form of a persistence assertion is as follows:

\medskip
\noindent
{\myfontcodesize
{\tt
:- persistent({\em PredSpec},{\em Connection})
}
}
\medskip

\noindent where {\tt {\em PredSpec}} is a predicate schema specification and the optional argument {\tt {\em Connection}} is an ODBC connection identifier.
{\tt {\em PredSpec}} can be either the pattern {\tt {\em PredName}}/{\tt {\em Arity}} or {\tt {\em PredName}}({\tt {\em Schema}}), where {\tt {\em Schema}} is the predicate schema, specified as: {\tt {\em ArgName$_1$}}:{\tt {\em Type$_1$}}, \ldots, {\tt {\em ArgName$_n$}}:{\tt {\em Type$_n$}}, where {\tt {\em ArgName$_i$}} are the argument names and {\tt {\em Type$_i$}} are their (optional) types for an $n$-ary predicate ($n>0$).
If a connection name is not provided, the name of the current open database is used, which must be an ODBC connection. 
An ODBC connection is identified by a name defined at the OS level, and opening a connection in DES means to make it the current database and that any relation (either a view or a table) defined in a DBMS is allowed as any other relation (predicate) in the deductive local database \verb+$des+.
A predicate can be made persistent only in one external database.
%

Any rule belonging to the definition of a predicate \verb+p+ which is being made persistent is expected, in general, to involve calls to other predicates (either directly or indirectly).
Each callee (such other called predicate) can be:
\begin{itemize}
\item An existing relation in the external database.
\item A persistent predicate loaded in the local database.
\item A persistent predicate not yet loaded in the local database.
\item A non-persistent predicate.
\end{itemize}
For the first two cases, besides making \verb+p+ persistent, nothing else is performed when processing its persistence assertion.
For the third case, a persistent predicate is automatically restored in the local database, i.e., it is made available to the deductive engine.
For the fourth case, each non-persistent predicate is automatically made persistent if types match; otherwise, an error is raised.
This is needed for the external database to be aware of a predicate only known by the deductive engine so far, as this database will be eventually involved in computing the meaning of \verb+p+.

\subsection{Implementing Persistence}
In general, a predicate is defined by extensional rules (i.e., facts) and intensional rules (including both head and body).
DES stores facts in a table and defines a view for the intensional rules.
For a predicate \verb+p+, a view with the same name as the predicate is created as the union of a table \verb+p_des_table+ (storing its extensional rules) and the equivalent SQL query for the remaining intensional rules.
This table is created resorting to the type information associated to \verb+p+.
So, given that a predicate \verb+p+ is composed of its extensional part {\tt {\em P}$_{ex}$} and its intensional part {\tt {\em P}$_{in}$},
each extensional rule in {\tt {\em P}$_{in}$} is mapped to a tuple in the table \verb+p_des_table+.
Let $\eval{\texttt{p}}{SQL}$ be the meaning of the view \verb+p+ in an SQL system,
and $\eval{\texttt{p}}{DL}$ be the meaning of the predicate \verb+p+ in the DES system, then:

\medskip
$\eval{\texttt{p}}{DL} = \eval{\texttt{p}}{SQL}$
\medskip

\noindent where the view \verb+p+ is defined by the following SQL query:

\medskip
{\myfontcodesize
\noindent {\tt
SELECT * FROM p\_des\_table UNION ALL \em{DL\_to\_SQL}(P$_{in}$)
}
}
\medskip

\noindent and {\tt {\em DL\_to\_SQL(P$_{in}$)}} is the function that translates a set of rules {\tt {\em P}$_{in}$} into an SQL query.
To this end, we have resorted to Draxler's Prolog to SQL compiler \cite{draxler1992powerful} (PL2SQL from now on), which is able to translate a Prolog goal into an SQL query.
Interfacing to this compiler is performed by the predicate {\tt translate(+{\em ProjectionTerm},+{\em PrologGoal},-{\em SQLQuery})}, where its arguments are, respectively, for: specifying the attributes that are to be retrieved from the database, defining the selection restrictions and join conditions, and representing the SQL query as a term.
So, a rule composed of a head {\tt {\em H}} and a body {\tt {\em B}} can be translated into an SQL query {\tt {\em S}} with the call {\tt translate({\em H},{\em B},{\em S})}.
Writing this as the function $dx\_translate(\texttt{\em R}_i)$, which is applied to a rule $\texttt{\em R}_i \equiv \texttt{\em H}_i :- \texttt{\em B}_i$ and returns its translated SQL query, and being {\tt {\em P$_{in}$} = \{$\texttt{\em R}_1, \ldots, \texttt{\em R}_n$\}}, then:

\medskip
{\myfontcodesize
\noindent
{\tt
{\em DL\_to\_SQL}({\em P}$_{in}$) =
{\tt $dx\_translate(\texttt{\em R}_1)$ UNION ALL \ldots UNION ALL $dx\_translate(\texttt{\em R}_n)$}
}
}
\medskip

PL2SQL is able to translate goals with conjunctions, disjunctions, negated goals, shared variables, arithmetic expressions in the built-in \verb+is+, and comparison operations, among others.
We have extended this compiler (PL2SQL$^+$ from now on) in order to deal with: Different, specific-DBMS-vendor code (including identifier delimiters and from-less SQL statements), the translation of facts, the mapping of some missing comparison operators, the inclusion of arithmetic functions to build expressions, and to reject both unsafe \cite{Ullman88} and recursive rules.
For instance, Access uses brackets as delimiters whereas MySQL uses back quotes.
Also, Oracle does not support from-less SQL statements and requires a reference to the table \verb+dual+, in contrast to other systems as PostgreSQL, which do not require it to deliver a one-tuple result (usually for evaluating expressions).
The predicate \verb+translate+ does not deal with true goals as they would involve a from-less SQL statement.
True goals are needed for translating facts, and so, we added support for this.
We have included arithmetic functions for the compilation of arithmetic expressions, including trigonometric functions (\verb+sin+, \verb+cos+, \ldots) and others (\verb+abs+, \ldots).
However, the support of such functions depends on whether the concrete SQL system supports them as well.
PL2SQL requires safe rules but it does not provide a check, so that we have included such a check to reject unsafe rules.
Recursive rules are not translated because not all DBMS's support recursive SQL statements (further DES releases might deal with specific code for recursive rules for particular DBMS's supporting recursion, as DB2 and SQL Server).
Figure \ref{fig:datalog-compiler} summarizes the syntax of valid inputs to PL2SQL$^+$ which are eventually represented as SQL statements.
Note that propositional predicates are not supported because relational databases require relations with arity greater than 0.

\begin{figure}
\begin{center}
\begin{tabular}{lll}
$fact$ & ::= & $p(c_1,\ldots,c_n)$ \\
$rule$ & ::= & $l ~ \texttt{:-} ~ l_1, \ldots, l_n$ \\
$l$    & ::= & $p(a_1,\ldots,a_n)$ \\
$l_i$  & ::= & $l ~|~ \texttt{not} ~ l ~|~ a_1 \lozenge a_2 ~|~ v ~ \texttt{is} ~~ e_1$ \\ 
$\lozenge$ & ::= & $\texttt{=} ~|~ \verb+\=+ ~|~ \texttt{<} ~|~ \texttt{=<} ~|~ \texttt{>} ~|~ \texttt{>=}$ \\
$e_i$  & ::= & $a ~|~ e_1 \blacklozenge e_2 ~|~ f(e_1,\ldots,e_n)$ \\
$\blacklozenge$ & ::= & $\texttt{+} ~|~ \texttt{-} ~|~ \texttt{*} ~|~ \texttt{/}$ \\ 
$f$ & ::= & $\texttt{sin} ~|~ \texttt{cos} ~|~ \texttt{abs} ~|~ \ldots$\\
\end{tabular}
\vspace*{2mm}

\begin{tabular}{cc}
$p$ is a predicate symbol. &
$c_i$ are constants, $i\geq 1$.\\
$l_i$ are literals, $i\geq 1$.&
$l$ is a term with depth 1.\\ 
$v$ is a variable. &
$a_i$ are either variables or constants, $i\geq 1$.\\
$e_i$ are arithmetic expressions. &
$rule$ is required to be safe and non recursive.\\
\end{tabular}
\vspace*{2mm}

\small{\em{True type symbols and pipes denote terminals and alternatives, respectively.}}
\end{center}
\caption{Valid Inputs to PL2SQL$^+$}
\label{fig:datalog-compiler}
\end{figure}

DES preprocesses Datalog rules before they can be eventually executed.
Preprocessing includes source-to-source transformations for translating several built-ins, including disjunction, outer joins, relational algebra division, top-N queries and others.
Rules sent to PL2SQL$^+$ are the result of these transformations, so that several built-ins that are not supported by PL2SQL can be processed by DES, as outer joins (left, right and full).
As well, there are other built-ins that PL2SQL$^+$ can deal with but which are not passed by DES up to now (as aggregates and grouping).

Non-valid rules for PL2SQL$^+$ but otherwise valid for DES are kept in the local database for their execution. 
%
In such a case, the deductive engine couples its own processing with the processing of the external database in the following way.
Let a predicate \verb+p+ be defined by a set of rules {\tt {\em S}} that can be externally processed and other set of rules {\tt {\em D}} that cannot.
Then, the meaning of \verb+p+ is computed as the union of the meanings of both sets of rules:

\medskip
$\eval{\texttt{p}}{} = \eval{\texttt{\em S}}{SQL} \cup \eval{\texttt{\em D}}{DL}$
\medskip

Rules in $\texttt{\em D}$ are therefore not included in {\tt {\em P}$_{in}$} in the call to {\tt {\em DL\_to\_SQL}} as described above, and they are otherwise stored as regular in-memory Datalog rules and processed by the deductive engine.
Thus, all the deductive computing power is preserved when either the external DBMS lacks some features as, e.g., recursion (e.g., MySQL and MS Access), or a predicate contains some non-valid rules for PL2SQL$^+$.

\subsection{An Example}
\label{sect:example}
As an example, let's consider the Datalog predicates \verb+ancestor+,  \verb+mother+, and \verb+parent+, the DBMS MySQL, and a table \verb+father+ already created and populated in this external database.

MySQL:
{\myfontcodesize
\begin{verbatim}
CREATE TABLE father(father VARCHAR(20),child VARCHAR(20));
INSERT INTO father VALUES('tom','amy');
...
\end{verbatim}
}

DES:
{\myfontcodesize
\begin{verbatim}
:-type(mother(mother:string,child:string)).
mother(grace,amy).
...

:-type(parent(parent:string,child:string)).
parent(X,Y) :- father(X,Y) ; mother(X,Y).

:-type(ancestor(ancestor:string,descendant:string)).
ancestor(X,Y) :- parent(X,Y).
ancestor(X,Y) :- parent(X,Z), ancestor(Z,Y).
\end{verbatim}
}

Then, if we submit the assertion \verb+:-persistent(ancestor/2)+ when the current opened database is MySQL, we get the following excerpt of the DES verbose output:

{\myfontcodesize
\begin{verbatim}
Warning: Recursive rule cannot be transferred to external database
  (kept in local database for its processing):
ancestor(X,Y) :- parent(X,Z), ancestor(Z,Y).
Info: Predicate mother/2 made persistent.
Info: Predicate parent/2 made persistent.
Info: Predicate ancestor/2 made persistent.
\end{verbatim}
}

Recalling Section \ref{sect:declaring-persistence}, declaring the persistence of \verb+ancestor/2+ involves to make persistent both \verb+mother/2+ and \verb+parent/2+ because, in particular, the first rule of \verb+ancestor/2+ includes a call to \verb+parent/2+, and the second call of \verb+parent/2+ is to \verb+mother/2+.
Even when \verb+parent/2+ includes a call to \verb+father/2+, the latter predicate is not made persistent because there exist the table \verb+father/2+ in the external database already.
The resulting views\footnote{They can be displayed, for instance, with the command {\tt /dbschema \$des}.} after processing the assertion are:

{\myfontcodesize
\begin{verbatim}
CREATE VIEW mother(mother,child) AS
  SELECT * FROM mother_des_table;

CREATE VIEW parent(parent,child) AS
  (SELECT * FROM parent_des_table) UNION ALL
  (SELECT rel1.mother,rel1.child FROM mother AS rel1) UNION ALL
  (SELECT rel1.father,rel1.child FROM father AS rel1);

CREATE VIEW ancestor(ancestor,descendant) AS
  (SELECT * FROM ancestor_des_table) UNION ALL
  (SELECT rel1.parent,rel1.child FROM parent AS rel1);
\end{verbatim}
}

Note that, on the one hand, and as a difference with other systems as $DLV^{DB}$, these views are not materialized.
On the other hand, DES allows to project such intensional rules to the external database by contrast to Ciao, which only project extensional rules.

Processing a top-level call either to \verb+father/2+, or \verb+mother/2+ or \verb+parent/2+ is computed by the external database.
However, a call to \verb+ancestor/2+ is processed both by the external database because of its first rule involving a call to \verb+parent+, and by the local deductive engine due to the local rule (the recursive one which cannot be processed by MySQL), as it will be explained in Section \ref{sect:intermixing}.

All intensional rules (both valid and non-valid inputs to PL2SQL$^+$) of a persistent predicate \verb+p+ are externally stored as metadata information in a table named \verb+p_des_metadata+ to allow to recover original rules when removing a persistence assertion (cf. Section \ref{subsect:restoring_persistence}).
For instance, the contents of this table for \verb+parent+ are \footnote{Note that as a result of DES preprocessing, the rule with the disjunction has been translated into two rules.}:
{\myfontcodesize
\begin{verbatim}
parent_des_metadata('parent(X,Y):-father(X,Y).').
parent_des_metadata('parent(X,Y):-mother(X,Y).').
\end{verbatim}
}
Whilst the contents of \verb+mother_des_table+ are its extensional rules (the facts  \verb+mother(grace,amy)+, \ldots), the contents of \verb+parent_des_table+ and \verb+ancestor_des_table+ are empty (unless a fact is asserted in any of the corresponding predicates).
Note that, as \verb+father+ is a table in the external database, if we {\em assert} a new tuple {\em t} for it, it will be only loaded in the local database, instead of externally stored if it was a persistent predicate\footnote{Of course, {\em inserting} a tuple in the external table will store it in the DBMS.}.
In both cases, the top-level query \verb+father(X,Y)+ would return the same tuples (either for the table or for the persistent predicate), but upon restoring persistence of \verb+ancestor/2+, the tuple {\em t} would not be restored for the table \verb+father+.


\subsection{Updating Persistent Predicates}

Updating a persistent predicate {\tt p} is possible with the commands \verb+/assert+ and \verb+/retract+, which allow to insert and delete a rule, respectively,
and their counterpart SQL statements \verb+INSERT+ and \verb+DELETE+, which allow to insert and delete, respectively, a batch of tuples (either extensionally or intensionally).
Implementing the update of the IDB part of a persistent predicate amounts to retrieve the current external view corresponding to the persistent predicate, drop it, and create a new one with the update.
The update of the EDB part (insert or delete a tuple) is simply performed to the external table with an appropriate SQL statement (\verb+INSERT INTO ...+ or \verb+DELETE FROM ...+).
Each update is tuple-by-tuple, even when batch updates via select statements are processed.
For each update, if constraint checking is enabled, any strong constraint defined at the deductive level is checked.

Note that the view update problem \cite{silberschatz2010database} is not an issue because our approach to insertions and deletions of tuples in a persistent predicate \verb+p+ amounts to modify the extensional part of \verb+p+, which is stored in the table \verb+p_des_table+.
This is a different approach to DBMS's where a relation defined by a view only consists of an intensional definition, so that trying to update a view involves updating the relations (other views and tables) it depends on, and this can be done is some situations but not in general.


\subsection{Restoring and Removing a Persistent Predicate}
\label{subsect:restoring_persistence}
Once a predicate \verb+p+ has been made persistent in a given session, the state of \verb+p+ can be restored in a next session (i.e., after the updates --assertions or retractions-- on \verb+p+)\footnote{Cf. transaction logic \cite{Bonner94anoverview} to model states in logic programming.}.
It is simply done by submitting again the same assertion as used to make \verb+p+ persistent for the first time.
Note, however, that if there exists any rule for \verb+p+ in the in-memory database already, it will not be removed but stored as persistent in the external database.
%
%
%
%

Also, a given predicate can be made non-persistent by dropping its assertion, as, e.g.:

{\myfontcodesize
\begin{verbatim}
DES> /drop_assertion :-persistent(p(a:int),mysql)
\end{verbatim}
}

This retrieves all the facts stored in the external database, stores them back in the in-memory database, removes them from the external database, and the original rules, as they were asserted (in its compiled Datalog form) are recovered from the table \verb+p_des_metadata+.
The view and tables for predicate \verb+p+ are dropped.

\subsection{Intermixing Query Solving}
\label{sect:intermixing}

As already introduced, persistence enables to couple external DBMS processing with deductive engine processing.
DES implements a top-down-driven, bottom-up fixpoint computation with tabling \cite{DBLP:conf/aplas/Saenz-PerezCG11}, which follows the ideas found in \cite{SD91,Dietrich87,TS86}.
This mechanism is implemented as described in \cite{Sae13a,sae13c-ictai13}.
In particular, the predicate {\tt solve\_goal} solves a goal (built-ins and user-defined predicates).
The following clause of this predicate is responsible of using program rules to solve a goal corresponding to a user predicate (where arguments which are not relevant for illustration purposes have been removed):

{\myfontcodesize
\begin{verbatim}
solve_goal(G) :- datalog((G:-B),_Source), solve(B).
\end{verbatim}
}

This predicate selects a program rule matching the goal via backtracking and solves the rule body as a call to the the predicate {\tt solve}.
Such program rules are loaded in the dynamic predicate {\tt datalog}.

In order to allow external relations to be used as user predicates, this dynamic predicate is overloaded with the following clause, which in turn calls {\tt datalog\_rdb}:

{\myfontcodesize
\begin{verbatim}
datalog(Rule,rdb(Connection)) :-
  datalog_rdb(Rule,rdb(Connection)).

datalog_rdb(R,Source) :-
  datalog_rdb_single_np(R,Source) ; % Single, non-persistent relation
  datalog_rdb_all_np(R,Source)    ; % All the non-persistent relations
  datalog_rdb_single_p(R,Source)  ; % Single, persistent predicate
  datalog_rdb_all_p(R,Source).      % All persistent predicates
\end{verbatim}
}

The predicate {\tt datalog\_rdb} identifies two possible sources: non-persistent and persistent predicates.
Also, it identifies whether a particular predicate is called or otherwise all predicates are requested.
In the last case, all external relations must be retrieved, and predicates {\tt datalog\_rdb\_all\_np} and {\tt datalog\_rdb\_all\_p} implement this via backtracking.
The (simplified) implementation of the predicate {\tt datalog\_rdb\_single\_p} (a single, concrete, persistent predicate) for an external ODBC connection {\tt Conn} is as follows:

{\myfontcodesize
\begin{verbatim}
datalog_rdb_single_p(R,RuleId,rdb(Conn)) :-
  my_persistent(Connection,TypedSchema),
  functor(TypedSchema,TableName,Arity),
  R =.. [Name|Columns],
  length(Columns,Arity),
  schema_to_colnames(TypedSchema,ColNames),
  sql_rdb_datasource(Conn,Name,ColNames,Columns,SQLstr),
  my_odbc_dql_query_fetch_row(Conn,SQLstr,Row),
  Row=..[_AnswerRel|Columns].
\end{verbatim}
}

The predicate {\tt sql\_rdb\_datasource} builds an SQL statement which returns rows for a relation under a connection matching the input column values ({\tt Columns} is the list of variables and/or constants for the query).
As an example, the query {\tt ancestor(A,amy)} for the example in Section \ref{sect:example} generates the following SQL statement (notice that the identifier delimiters in this DBMS do not follow standard SQL):

{\myfontcodesize
\begin{verbatim}
SELECT * FROM `ancestor` WHERE `descendant`='amy'
\end{verbatim}
}

The predicate {\tt my\_odbc\_dql\_query\_fetch\_row} returns rows, one-by-one, via backtracking for this SQL statement.
Note that, for this simple example, row filtering is performed by the external engine.

Recall that this persistent predicate consists of two program rules:

{\myfontcodesize
\begin{verbatim}
ancestor(X,Y) :- parent(X,Y).
ancestor(X,Y) :- parent(X,Z), ancestor(Z,Y).
\end{verbatim}
}

The first one was loaded in the external database as the view:

{\myfontcodesize
\begin{verbatim}
CREATE VIEW ancestor(ancestor,descendant) AS
  (SELECT * FROM ancestor_des_table) UNION ALL
  (SELECT rel1.parent,rel1.child FROM parent AS rel1);
\end{verbatim}
}

\noindent and the second one was loaded in the local deductive database, as the dynamic clause:

{\myfontcodesize
\begin{verbatim}
datalog((ancestor(X,Y) :- parent(X,Z), ancestor(Z,Y)),source)
\end{verbatim}
}

So, the fixpoint mechanism uses in the call to \verb+datalog+ both the non-recursive rule from the external database via {\tt datalog\_rdb\_single\_p}, and the recursive rule via the dynamic clause.
Concluding, the predicate {\tt datalog} provides to the deductive query solving not only the rules which are local, but also the rules which are externally stored and processed, retrieved via the predicate {\tt datalog\_rdb}, therefore enabling intermixed query solving.

%

\subsection{Fixpoint Optimizations}
\label{subsect:optimization}

We list some already implemented optimizations which are key to avoid retrieving the same tuple from the external database several times due to fixpoint iterations.
They can be independently enabled and disabled with commands to test their impact on performance.

\begin{itemize}

\item {\bf Complete Computations}.
Each call during the computation of a stratum is remembered in addition to its outcome (in the answer table).
Even when the calls are removed in each fixpoint iteration, most general ones do persist as a collateral data structure to be used for saving computations should any of them is called again during either computing a higher stratum or a subsequent query solving.
If a call is marked as a completed computation, it is not even tried if called again.
This means the following two points:
1) During the computation of the memo function, calls already computed are not tried to be solved again, and only the entries in the memo table are returned.
2) Moreover, computing the memo function is completely avoided if a subsuming already-computed call can be found.
In the first case, that saves solving goals in computing the memo function.
In the second case, that completely saves fixpoint computation.

\item {\bf Extensional Predicates}.
There is no need to iteratively compute extensional predicates and, therefore, no fixpoint computation is needed for them.
They are known from the predicate dependency graph simply because they occur in the graph without incoming arcs.
For them, a linear fetching is enough to derive their meanings.


\item {\bf Non-Recursive Predicates}.
Memoization comes at the cost of maintaining a cache which can be wasteful in some cases.
A top-level goal involving non-recursive predicates are computed by only caching the top-level goal, avoiding memorizing dependent calls.
This allows a fast solving by looking for all the answers of the goal, and finally storing the results in the memo table.

\end{itemize}

%% file: 05.tex
\section{Applications}
\label{sect:applications}

This section lists several applications derived from supporting persistence in DES as it includes some features which are not available in external DBMS's, such as hypothetical queries, extended recursion, and intermixed query solving.

\subsection{Database Interoperability}
Persistence allows for database interoperability as each persistent predicate is mapped to an ODBC connection and several connections can be opened simultaneously.
First scenario is for a persistent predicate \verb+p+ in a given connection and opening another connection from another database.
Then, both the predicate \verb+p+ and the relations defined in the latter connection are visible for the deductive database.
This is in contrast to other systems (as, e.g., $DLV^{DB}$) that need to explicitly state what relations from the external database are visible.
Here, no extra effort is needed.
Second scenario is for several persistent predicates which are mapped to different connections.
As they are visible for the deductive engine, all of them can be part of a query solved by the deductive engine.
Recall that any external view will be still processed by the external DBMS.


%
%
%
%

\subsection{Extending DBMS Expressivity}
\label{sect:extending-expressivity}
The more expressive SQL and Datalog languages as provided by DES can improve the expressiveness of the external database when acting as a front-end.
For instance, let's consider MySQL, which does not support recursive queries up to its current version 5.6.
The following predicate can be made persistent in this DBMS even when it is recursive:

{\myfontcodesize
\begin{verbatim}
DES> :-persistent(path(a:int,b:int),mysql)
DES> /assert path(1,2)
DES> /assert path(2,3)
DES> /assert path(X,Y):-path(X,Z),path(Z,Y)
Warning: Recursive rule cannot be transferred to external database (kept
    in local database for its processing):
path(X,Y) :- path(X,Z), path(Z,Y).
DES> path(X,Y)
{ path(1,2), path(1,3), path(2,3) }
\end{verbatim}
}

Here, non-recursive rules are stored in the external database whereas the recursive one is kept in the local database.
External rules are processed by MySQL and local rules by the deductive engine.
Though the recursive rule is not externally processed, it is externally stored as metadata, therefore ensuring its persistence between sessions.
%

In addition to Datalog, DES includes support for SQL for its local deductive database.
To this end, on the one hand, SQL data definition statements are executed and metadata (as the name and type of table fields) is stored as assertions.
On the other hand, SQL queries are translated into Datalog and executed by the deductive engine.
The supported SQL dialect includes features which are not found in current DBMS's, as non-linear recursive queries, hypothetical views and queries, and the relational algebra division operator.
Therefore, DES is able to compute more queries than a DBMS:
For instance, neither MS SQL Server nor IBM DB2 allow cycles in a path without compromising termination.
Also, recursive and stratifiable SQL queries do not fully allow \verb+EXCEPT+ such as in MS SQL Server and IBM DB2.
Another limitation is linear recursion: The rules in the last example above cannot be expressed in any DBMS as there are several recursive calls.
To name another, \verb+UNION ALL+ is enforced in those SQL's, so that just \verb+UNION+ (discarding duplicates) is not allowed.
For instance, the following recursive query is rejected in any current commercial DBMS, but accepted by DES:

{\myfontcodesize
\begin{verbatim}
DES> CREATE TABLE edge(a int, b int);
DES> INSERT INTO edge VALUES (1,2),(2,3),(1,3);
DES> :-persistent(edge/2,mysql).
DES> :-persistent(path(a:int,b:int),mysql).
DES> WITH RECURSIVE path(a, b) AS
  SELECT * FROM edge
  UNION --Discard duplicates (ALL not required)
  SELECT p1.a,p2.b FROM path p1, path p2 WHERE p1.b=p2.a
SELECT * FROM path;
Warning: Recursive rule cannot be transferred to external database
  (kept in local database for its processing):
path_2_1(A,B) :- path(A,C), path(C,B).
answer(path.a:number(integer), path.b:number(integer)) ->
{ answer(1,2), answer(1,3), answer(2,3) }
\end{verbatim}
}

In this example, \verb+edge+ becomes a Datalog typed (and populated) relation because it is defined with the DES SQL dialect in the local deductive database, and it has been made persistent, as well as \verb+path+ (which is also typed because of the persistence assertion, but not populated).
The \verb+WITH+ statement allows to declare temporary relations.
In this case, the result of the compilation of the SQL query definition of \verb+path+  are temporary Datalog rules which are added to the persistent predicate \verb+path+ (note that the recursive part is not transferred to the external database):

{\myfontcodesize
\begin{verbatim}
path(A,B) :- distinct(path_2_1(A,B)).
path_2_1(A,B) :- edge(A,B).
path_2_1(A,B) :- path(A,C), path(C,B).
\end{verbatim}
}

\noindent and the SQL query \verb+SELECT * FROM path+ is compiled to:

{\myfontcodesize
\begin{verbatim}
answer(A,B) :- path(A,B).
\end{verbatim}
}

After executing the goal \verb+answer(A,B)+ for solving the SQL query, the temporary Datalog rules are removed.
Adding {\tt ALL} to {\tt UNION} to the same query for keeping duplicates makes to include the tuple {\tt answer(1,3)} twice in the result.




\subsection{Business Intelligence}
\label{sect:business-intelligence}
Business intelligence refers to systems which provide decision support \cite{Watson:2007:CSB:1300761.1301970} by using data integration, data warehousing, analytic processing and other techniques.
In particular, one of these techniques refer to ``what-if" applications.
DES also supports a novel SQL feature: Hypothetical SQL queries.
Such queries are useful, for instance, in decision support systems as they allow to submit a query by assuming some knowledge which is not in the database.
Such knowledge can be either new data assumed for relations (both tables and views) and also new production rules.
For example, and following the above system session, the tuple \verb+(3,1)+ is assumed to be in the relation \verb+path+, and then this relation is queried:

{\myfontcodesize
\begin{verbatim}
DES> ASSUME SELECT 3,1 IN path(a,b) SELECT * FROM path;
answer(path.a:number(integer),path.b:number(integer)) ->
{ answer(1,1), answer(1,2), answer(1,3), answer(2,1), answer(2,2),
  answer(2,3), answer(3,1), answer(3,2), answer(3,3) }
\end{verbatim}
}

As an example of adding a production rule, let's suppose a relation {\tt flight} and a view {\tt connect} for locations connected by direct flights:

{\myfontcodesize
\begin{verbatim}
DES> CREATE TABLE flight(ori STRING, dest STRING, duration INT);
DES> INSERT INTO flight VALUES ('Madrid','Paris',90),
       ('Paris','Oslo',100), ('Madrid','London',110);
DES> CREATE VIEW connect(ori,dest) AS SELECT ori,dest FROM flight;
DES> :-persistent(connect/2,access) -- This also makes 'flight' persistent
DES> SELECT * FROM connect;
answer(connect.ori:string(real),connect.dest:string(real)) ->
{ answer('Madrid','London'), answer('Paris','Oslo'),
  answer('Madrid','Paris') }
\end{verbatim}
}
Then, if we assume that connections are allowed with transits, we can submit the following hypothetical query (where the assumed SQL statement is recursive):

{\myfontcodesize
\begin{verbatim}
DES> ASSUME
      (SELECT flight.ori,connect.dest
       FROM flight,connect
       WHERE flight.dest = connect.ori)
     IN
       connect(ori,dest)
     SELECT * FROM connect;
answer(connect.ori:string(real),connect.dest:string(real)) ->
{ answer('Madrid','London'),answer('Madrid','Oslo'),
  answer('Madrid','Paris'), answer('Paris','Oslo')}
\end{verbatim}
}

Also, several assumptions for different relations can be defined in the same query.

\subsection{Migrating Data}
\label{sect:migration}

Once a predicate has been made persistent in a given connection, dropping its persistent assertion retrieves all data and schema from the external database into the in-memory Prolog database.
A successive persistent assertion for the same predicate in a different connection dumps it to the new external database.
These two steps, therefore, implement the migration from one database to another, which can be of different vendors.
For instance, let's consider the following session, which dumps data from MS Access to MySQL:

{\myfontcodesize
\begin{verbatim}
DES> :-persistent(p(a:int),access)
DES> /drop_assertion :-persistent(p(a:int),access)
DES> :-persistent(p(a:int),mysql)
\end{verbatim}
}


%% file: 06.tex
\section{Performance}
\label{sect:performance}

In this section we analyze how queries involving persistent predicates perform w.r.t. native SQL queries, and the overhead caused by persistence w.r.t. the in-memory (Prolog-implemented) database.

As relational database systems, three widely-used systems have been chosen with a default configuration: The non-active desktop database MS Access (version 2003 with ODBC driver 4.00.6305.00), the mid-range, open-source Oracle MySQL (version 5.5.9 with ODBC driver 5.01.08.00), and the full-edged, commercial IBM DB2 (version 10.1.0 with ODBC driver 10.01.00.872).
All times are given in milliseconds and have been run on an Intel Core2 Quad CPU at 2.4GHz and 3GB RAM, running Windows XP 32bit SP3.
Each test has been run 10 times, the maximum and the minimum numbers have been discarded, and then the average has been computed.
Also, as Access quickly fragments the single file it uses for persistence, and this heavily impacts performance, each running of the benchmarks in this system is preceded by a defragmentation (though, the time for performing this has not been included in the numbers).
All optimizations, as listed in Section \ref{sect:intermixing}, are enabled.

Some results are collected in Table \ref{table:performance1}.
The tests consist of, first, inserting 1,000 tuples in a relation with a numeric field (columns with heading {\em Insert$_n$} and {\em Insert$_p$}, for native queries and persistent predicates, respectively).
The Datalog commands are {\tt /assert t({\em i})} and the SQL update queries are {\tt INSERT INTO t VALUES({\em i})} (1 $\leq$ {\tt {\em i}} $\leq$ 1,000).

\begin{table}[b!]
\begin{center}
\resizebox{\columnwidth}{!}{
\begin{tabular}{|c|r@{}@{}rr@{}@{}rr@{}@{}r|r@{}@{}rr@{}@{}rr@{}@{}r|}
\cline{1-7}
{\em System} & \multicolumn{2}{|c}{\em Insert$_n$} & \multicolumn{2}{c}{\em Select$_n$} & \multicolumn{2}{c|}{\em Join$_n$} &\\
\hline
DES 3.2  &   359  &        &   773  &      &  3,627 &       & \multicolumn{2}{c}{\em Insert$_p$} & \multicolumn{2}{c}{\em Select$_p$} & \multicolumn{2}{c|}{\em Join$_p$}\\
\cline{8-13}
Access   &   439  &  (1.22) & 1,014 & (1.31) &  7,303 & (2.01) &  1,102 &  (3.07$\diamond$2.51) & 2,138 & (2.77$\diamond$2.11) & 17,270 & (4.76$\diamond$2.36)\\
MySQL    & 9,950  & (27.72) & 1,160 & (1.50) & 13,183 & (3.63) & 10,279 & (28.63$\diamond$1.03) & 2,364 & (3.06$\diamond$2.04) & 22,305 & (6.15$\diamond$1.69)\\
DB2      & 1,264  &  (3.52) & 1,018 & (1.32) &  9,057 & (2.50) &  1,869 &  (5.21$\diamond$1.48) & 2,260 & (2.92$\diamond$2.22) & 18,637 & (5.14$\diamond$2.06)\\
\hline
\end{tabular}
}
\end{center}
\caption{Results for in-memory DES, DBMS's and Persistent Predicates}
\label{table:performance1}
\end{table}

Then, 1,000 select queries are issued (columns {\em Select$_n$} and {\em Select$_p$}).
The $i$-th select query asks for the $i$-th value stored in the table, so that all values are requested by independent queries.
The Datalog queries are {\tt t({\em i})} and the SQL select queries are {\tt SELECT a FROM t WHERE a={\em i}} (1 $\leq$ {\tt {\em i}} $\leq$ 1,000).

Next, a single query which computes an autojoin is submitted (columns {\em Join$_n$} and {\em Join$_p$}) which yields one million tuples in the result set.
The Datalog queries are {\tt t(X),t(Y)} and the SQL select queries are {\tt SELECT * FROM t AS t1,t AS t2}.

First line below headings of this table collects the results of the in-memory deductive database DES (Datalog commands and queries), with no persistence.
The next three lines in the block with subscripts $n$ in the headings (referred to as 'block $n$' from now on) show the results for the native queries in each DBMS (SQL {\tt INSERT} and {\tt SELECT} queries).
The three lines in the block with subscripts $p$ in the headings (referred to as 'block $p$' from now on) show the results for the Datalog commands (\verb+/assert+ {\tt t({\em i})}) and queries ({\tt t({\em i})} and \verb+t(X),t(Y)+) when the relation {\tt t} has been made persistent in each external DBMS.

Then, this table allows, first, to compare the in-memory, state-less system DES w.r.t. the relational, durable DBMS's (ratio values enclosed between parentheses in block $n$ as the time for each DBMS divided by the time for DES).
Second, to examine the overhead of persistence by confronting the results in the line DES and the results in the block $p$ for each DBMS (first ratio value enclosed between parentheses in the table as the time for DES divided by the time for each DBMS).
And, third, to compare the results of DES as a persistent database w.r.t. each DBMS for dealing with the same actions (inserting and retrieving data), by confronting the results in block $p$ and block $n$ for each DBMS (second ratio value enclosed between parentheses in the table as the time for the time in block $p$ divided by the corresponding time in block $n$).

For the {\tt SELECT} queries, we focus on retrieving to the main memory the results but without actually displaying it in order to elide the display time.
For the deductive database, this means that each tuple in the result is computed and stored in the answer table but it is not displayed.
For the relational databases, this means that a single ODBC cursor connection is used for a single query and each tuple in its result is retrieved to main memory, but not displayed.



With respect to the native queries (focusing at block $n$) a first observation is that insertions (column {\em Insert$_n$}) in the in-memory deductive database are, as expected, faster than for DBMS's.
However, Access is very fast as it is more oriented towards a file system (it is not an active database) and its time is comparable to that of DES (Access is only 22\% slower).
Another observation is that MySQL takes much more time for updates than DB2 (9,950/1,264 $\approx$ 8 times slower) and Access (9,950/439 $\approx$ 22.6 times slower), but it performs close to them for the batch of 1,000 select queries.
(This behaviour can also be observed in queries to persistent data.)
A third observation is that computations for select operations (columns {\em Select}$_n$ and {\em Join}$_n$) in the in-memory deductive database are faster than in DBMS's.
While for 1,000 queries in Datalog (column {\em Select}$_n$) there is a speed-up of up to 1.50, in the single query (column {\em Join}$_n$) this grows up to 3.63 (both for MySQL).

Queries to persistent data (focusing at block $p$) show two factors: 1) The performance of queries involving persistence w.r.t. their counterpart native SQL queries, and 2) The overhead caused by persistence in the deductive system for the different DBMS's.
%
%
With respect to factor 1, by comparing native queries to queries to persistent data, we observe that the cost for inserting tuples by using a persistent predicate w.r.t. a native SQL {\tt INSERT} statement ranges from a negligible ratio of 1.03 (MySQL) to 2.51 (Access).
Also, the overhead for computing 1,000 queries with a Datalog query on a persistent predicate w.r.t. its counterpart native SQL select statement, is around 2 times for all DBMS's.
And for the autojoin, the ratio ranges from 1.69 for MySQL to 2.36 for Access.
With respect to factor 2, insertions require a ratio ranging from 3.07 to 5.21 for Access and DB2, respectively, whereas for DB2 a huge ratio of 28.63 is found.
Managing individual insert statements via cursor connections is hard in this case.
However, the overhead comes from the connection itself as the code to access the different external databases is the same.
The select queries perform quite homogeneously with ratios from 2.76 to 3.06, in accordance to factor 1.
Last, for the autojoin, the ratio ranges from 4.76 to 6.15.

Finally, Table \ref{table:performance2} shows the cost for creating and removing persistence for each DBMS.
The column {\em Create} shows the time for creating a persistent predicate where its 1,000 tuples are in the in-memory database.
This amounts to store each in-memory tuple in the external database, so that numbers are similar to that of the column {\em Insert}$_n$.
Dropping the persistent assertion, as shown in column {\em Drop}, takes a small time.
Recall that this operation also retrieve the 1,000 tuples to the in-memory database.
The difference between the cost of creating and dropping the assertion lies in that the former submits 1,000 SQL queries while the latter submits a single SQL query.
Thus, the cost of opening and closing cursor connections is therefore noticeable.

\begin{table}[hb]
\begin{center}
\begin{tabular}{|c|rr|}
\hline
{\em DBMS} & {\em Create} & {\em Drop}\\
\hline
Access   &  1,256 &  31\\
MySQL    & 10,523 &  74\\
DB2      &  1,926 & 172\\
\hline
\end{tabular}
\end{center}
\caption{Creating and Removing Persistence}
\label{table:performance2}
\end{table}

%% file: 07.tex
\section{Conclusions}
\label{sect:conclusions}
This paper has shown how persistence is supported by a tabled-based deductive system.
This work includes extended language features that might be amenable to try even projected to such external databases.
Although this system was targeted at teaching and not to performance, some numbers have been taken to assess its applicability.
When comparing the times taken by the queries relating persistent predicates w.r.t. their counterpart native SQL queries, ratios from 1.03 up to 2.51 are got, which overall shows the overhead of using the deductive persistent system w.r.t. the SQL systems.
When comparing the times taken by the queries relating persistent predicates w.r.t. their counterpart in-memory queries, higher ratios have been found, from 2.76 up to 6.15, and an extreme case of 28.63 due to the costly insertions through the ODBC bridge.
These results suggest that the cost of persistence might be worthwhile depending on the DBMS and the application.

Differences between this system and others can be highlighted, besides those which were already noted in the introduction and along the paper.
For instance, predicates in DLV$^{DB}$ are translated into materialized relations, i.e., a predicate is mapped to a table and the predicate extension 
is inserted in this table, which opens up the view maintenance problem.
Ciao Prolog is only able to make the extensional part of a predicate to persist, disabling the possibility of surrogating the solving of views for intensional rules.
MyYapDB (for *unixes) is not understood as implementing persistence, instead, it allows to connect to the external MySQL DBMS, making external relations available to YAP as if they were usual predicates.
This is similar to what DES does simply by opening an ODBC connection, which automatically makes visible all the external relations (not only in MySQL but for any other DBMS and OS).
LDL++ was retired in favor of DeAL, and currently there is no information about its connection to external databases, though in \cite{ldl-2003} such a connection was very briefly described for the former.

As for future work, built-ins supported by the compiler \cite{draxler1992powerful} but not passed by DES can be included in forthcoming releases.
Also, query clustering can be useful (cf. \cite{DBLP:conf/padl/CorreasGCCH04}), i.e., identifying those complex subgoals that can be mapped to a single SQL query, therefore improving the results for queries as the autojoin, by reducing the number of cursors.
Rules with linear recursive queries supported by the external DBMS can be allowed to be projected.
Since the deductive engine is not as efficient as others \cite{DBLP:journals/tplp/SwiftW12}, it can be improved or replaced with an existing one but upgraded to deal with extra features (as nulls and duplicates).
Finally, the current implementation has been tested for several DBMS's, including Access, SQL Server, MySQL, and DB2.
Although the connection to such external databases is via the ODBC bridge which presents a common interface to SQL, some tweaks depending on the particular SQL dialect should be made in order to cope with other DBMS's.

%% file: 08.tex
\section*{Acknowledgements}
This work has been partially supported by the Spanish projects CAVI-ART (TIN2013-44742-C4-3-R), STAMP (TIN2008-06622-C03-01), Prometidos-CM (S2009TIC-1465), GPD (UCM-BSCH-GR35/10-A-910502), and the Department Ingenier\'{\i}a del Software e Inteligencia Artificial at University Complutense of Madrid.
Also thanks to the anonymous referees who helped in improving this paper and the system DES. 